\documentclass{elsart}

\usepackage{amssymb}

\journal{J. Math. Anal. Applcs.}

\begin{document}
\begin{frontmatter}
\title{A class of identities relating Whittaker and Bessel functions}
\author{James Lucietti}
\address{ \small Department of Applied Mathematics and Theoretical Physics,\\ University of Cambridge, UK}
\ead{J.Lucietti@damtp.cam.ac.uk}

\begin{abstract}
Identities between Whittaker and modified Bessel functions are
derived for particular complex orders. Certain polynomials appear in
such identities, which satisfy a fourth order differential equation
(not of hypergeometric type), and they themselves can be expressed
as particular linear combinations of products of modified Bessel and
confluent hypergeometric functions.
\end{abstract}
\end{frontmatter}

\section{Introduction}
A class of identities is derived which express Whittaker functions
$W_{N,ik}(2x)$ in terms of modified Bessel functions of the second
kind, where $k$ is real, $N$ is integer or half-integer. In this
paper we concentrate on the case where $N=n+1/2$ where $n$ is a
natural number. More explicitly, we will find that

\begin{equation}
W_{n+1/2,ik}(2x) = x\Lambda^{k}_n(x) K_{1/2+ik}(x) +x{\Lambda^{k}_n}^*(x)K_{1/2-ik}(x)
\end{equation}
where $\Lambda^{k}_n(x)$ is a polynomial of degree $n$. These
polynomials reduce to Laguerre polynomials when $k=0$ and we will be
able to express them as a particular linear combination of products
of modified Bessel and confluent hypergeometric functions.

 We should
note that the $n=0$ case of this identity was noticed in the
solution of a ``physical'' problem; namely the energy eigenfunctions
of supersymmetric quantum mechanics with an exponential
potential~\cite{superliouville,me}.

\section{Proof of identities}
We begin by writing down Whittaker's differential equation~\cite{ww},

\begin{equation}
L(y) \equiv y''(x) + \left( -1 + \frac{2n+1}{x} + \frac{\frac{1}{4} +k^2}{x^2}
\right)y(x)=0,
\end{equation}
which possesses $W_{n+1/2,ik}(2x)$ as a solution. Our strategy will
be to show that $\lambda^{k}_n(x) K_{1/2+ik}(x)
+{\lambda^{k}_n}^*(x)K_{1/2-ik}(x)$ satisfies this differential
equation for some polynomial $\lambda^{k}_n(x)$ and determine the
polynomial as a byproduct. Then studying the asymptotics will
complete the proof.

Substituting $\lambda^{k}_n(x) K_{1/2+ik}(x)
+{\lambda^{k}_n}^*(x)K_{1/2-ik}(x)$ into the differential equation
we arrive at

\begin{eqnarray}
&& L(\lambda^{k}_n K_{1/2+ik} +\textrm{c.c.})=
 2{\lambda_n^k}' K'_{1/2+ik} +\lambda^k_n {K''}_{1/2+ik} + \\
&& \left( {\lambda^k_n}''
-\lambda^{k}_n + \frac{2n+1}{x}\lambda^{k}_n +  \frac{\frac{1}{4} +k^2}{x^2}\lambda^{k}_n \right)
K_{1/2+ik} + \textrm{c.c.},\nonumber
\end{eqnarray}
where $+ \textrm{c.c.}$ means add the complex conjugate of the preceding terms.
We can eliminate the second derivatives ${K''}_{1/2+ik}(x)$ by using
Bessel's equation,

\begin{eqnarray}
{K''}_{1/2+ik} + \frac{1}{x}K'_{1/2+ik} - \left(
1+\frac{(1/2+ik)^2}{x^2} \right)K_{1/2+ik} = 0,
\end{eqnarray}
and this gives,
\begin{eqnarray}
&&\left(2{\lambda_n^k}' - \frac{\lambda^{k}_n}{x} \right) K'_{1/2+ik} + \left(
\frac{1/2+ik}{x^2}\lambda^{k}_n +\frac{1+2n}{x}\lambda^{k}_n + {\lambda^k_n}'' \right)K_{1/2+ik} +
\textrm{c.c.}  \nonumber \\ &&=L(\lambda^{k}_n K_{1/2+ik} +\textrm{c.c.}). \
\end{eqnarray}
Now, we can eliminate the first derivatives $K'_{1/2+ik}(x)$ using
the identities~\cite{gr}

\begin{eqnarray}
\label{first}
 &&xK'_{\nu}(x) \pm \nu K_{\nu}(x) = -x K_{\nu\mp 1}(x), \\
 &&K_{\nu}(x)=K_{-\nu}(x)
\end{eqnarray}
which lead to

\begin{eqnarray}
\left[ {\lambda^k_n}'' -\frac{1+2ik}{x}{\lambda_n^k}' + \left( \frac{1+2ik}{x^2}
+\frac{1+2n}{x}\right)\lambda^{k}_n \right] K_{1/2+ik} +\\ \nonumber \left( \frac{\lambda^{k}_n}{x} -
2{\lambda_n^k}' \right) K_{1/2-ik} + \textrm{c.c.} =  L(\lambda^{k}_n K_{1/2+ik} +\textrm{c.c.}).
\end{eqnarray}
The complex conjugate term is not independent. Since
$K_{1/2+ik}^*(x)= K_{1/2-ik}(x)$ we can rewrite the whole expression
above as $(...)K_{1/2+ik}+ \textrm{c.c.} $ and this can be made to
vanish if the coefficient of $K_{1/2+ik}(x)$ is made to vanish; this
condition corresponds to

\begin{eqnarray}
{\lambda^k_n}'' - \frac{1+2ik}{x} {\lambda_n^k}' + \left( \frac{1+2ik}{x^2}
+\frac{1+2n}{x} \right)\lambda^{k}_n + \left( \frac{{\lambda^{k}_n}^*}{x} - 2{{\lambda^k_n}'}^*
\right)=0.
\end{eqnarray}
Of course one can consider the complex conjugate version of this
differential equation, and then we can view them as two linear
second-order coupled differential equations for $\lambda^{k}_n$ and
${\lambda^{k}_n}^*$. This will imply that $\lambda^{k}_n$ satisfies
a fourth-order linear ODE. If we substitute $\lambda^{k}_n(x)=
\sum_{m=0}^{n+1} a_m^{(n)}x^m$, we can derive a recurrence relation
for the coefficients $a^{(n)}_m$. We find that

\begin{eqnarray}
&& a^{(n)}_0=0, \\
&& m(m+1)(2m-1)(m+2ik)(m-1-2ik)a^{(n)}_{m+2} + \\ \nonumber
&& (1+2n)m(3m^2+m-2ik)a^{(n)}_{m+1} - \\
&& 4(1+2m)(n+m)(1+n-m)a^{(n)}_{m} = 0, \quad 1 \leq m \leq n-1. \nonumber
\end{eqnarray}
This is a rather complicated recurrence relation,; in particular it
does not generate a hypergeometric series; however given any two
members of the sequence it clearly determines the rest. Thus, now we
proceed to determine two of the coefficients using the asymptotics
of the functions. Note that it is actually convenient to consider
the recurrence relation one gets directly from the differential
equation above. This is,

\begin{eqnarray}
&&m(m-2ik)a^{(n)}_{m+1} + (1+2n)a^{(n)}_{m} +
(1-2m){a}^{(n)*}_{m} = 0, \quad 1 \leq m \leq n \\
&&a^{(n)*}_{n+1}- a^{(n)}_{n+1} = 0
\end{eqnarray}
and from this one gets to the second order recurrence relation above
by eliminating ${a}^{(n)*}_m$. It is known~\cite{ww} that as $x \to \infty$,

\begin{eqnarray}
&&W_{n+1/2,ik}(2x) \sim (2x)^{n+1/2} e^{-x},\\
&&K_{\nu}(x) \sim \sqrt{\frac{\pi}{2x}}e^{-x},
\end{eqnarray}
which allows us to deduce

\begin{eqnarray}
\lambda^{k}_n(x) + {\lambda^{k}_n}^*(x) \sim \frac{2^{n+1}}{\sqrt{\pi}} x^{n+1},
\end{eqnarray}
and this tells us that

\begin{equation}
a^{(n)}_{n+1} + a^{(n)*}_{n+1}  = \frac{2^{n+1}}{\sqrt{\pi}},
\end{equation}
providing us with enough information to solve for $a^{(n)}_{n+1}$
and we find

\begin{equation}
a^{(n)}_{n+1} = \frac{2^{n}}{\sqrt{\pi}}.
\end{equation}
Now we turn to the asymptotics for small $x$. We will
need~\cite{ww,gr}

\begin{eqnarray}
\label{Wasymptotics} &&W_{n+1/2,ik}(2x) \sim
\frac{\Gamma(-2ik)}{\Gamma(-ik-n)} (2x)^{1/2+ik} + \textrm{c.c.} \qquad \textrm{as} \;x \to 0\\
&&K_{\nu}(x)= \sqrt{\frac{\pi}{2x}}W_{0,\nu}(2x),
\end{eqnarray}
from which we can derive,

\begin{eqnarray}
\lambda^{k}_n(x)K_{1/2+ik}(x) + \textrm{c.c} \sim a^{(n)*}_1 2^{ik}
\sqrt{\frac{\pi}{2}} \frac{\Gamma(1-2ik)}{\Gamma(1-ik)} x^{1/2+ik} + \textrm{c.c.}
\end{eqnarray}
and upon comparison to (\ref{Wasymptotics}), obtain

\begin{equation}
a^{(n)}_1 = \frac{1}{\sqrt{\pi}} \frac{\Gamma(-ik)}{\Gamma(-n-ik)}=
\frac{(-1)^n}{\sqrt{\pi}} (1+ik)_n.
\end{equation}
Thus we have derived the first and last coefficient in the polynomial,
which together with the recurrence relation serve to define $\lambda^{k}_n(x)$
uniquely. Note that the identity is now actually proved as we have
shown that $\lambda^{k}_n(x)K_{1/2+ik}(x) + \textrm{c.c}$ satisfies the same
differential equation as $W_{1/2+n,ik}(2x)$ and possesses the same
asymptotics and thus they must be the same function.

\section{The polynomials $\Lambda^k_n(x)$}
Let us consider the special case $k=0$ for which the identity
reduces to a well known one. It is clear that in this case the
polynomials are real, since both $a^{(n)}_m$ and $a^{(n)*}_m$
satisfy the same recurrence relation and boundary conditions. Then
we see that the polynomials actually satisfy a second order
differential equation and if we let $\lambda^{0}_n(x)= xy_n(2x)$ we
find

\begin{equation}
z{y''}_n(z) + (1-z){y'}_n(z) +ny_n(z)=0,
\end{equation}
and thus we see that $y_n(z)= c_n L_n(z)$ where $L_n(z)$ are the
Laguerre polynomials~\cite{gr}. Once again, asymptotics can be used
to determine the proportionality constants $c_n$. Therefore, since
$L_n(z) \sim 1$ as $z \to 0$, and $\lambda^0_n(x) \sim (-1)^n n! x /
\sqrt{\pi}$  as $x\to 0$, we see that $c_n=  (-1)^n n! / \sqrt{\pi}$
and thus

\begin{equation}
\lambda^0_n(x)= \frac{(-1)^n n!}{\sqrt{\pi}}\,x \,L_n(2x).
\end{equation}

Now, we give the fourth order equation that the polynomials satisfy.
First introduce $ \lambda^k_n(x)=x\Lambda^k_n(x)$; then we have

\begin{equation}
\label{coupled}
x{\Lambda^k_n}'' + (1-2ik){\Lambda^k_n}' +(1+2n)\Lambda^k_n
-2x{{\Lambda^k_n}'}^*- {\Lambda^k_n}^*=0,
\end{equation}
as our second order equation, and after some work one can eliminate
${\Lambda^k_n}^*$ to get the rather unsightly answer

\begin{eqnarray}
&&a_1(x){\Lambda^k_n}'''' +a_2(x){\Lambda^k_n}'''
+a_3(x){\Lambda^k_n}''+ a_4(x){\Lambda^k_n}' +a_5(x)\Lambda^k_n =0, \\
&&a_1(x)=x^2[1-4ik+4x(1+2n)], \nonumber \\
&&a_2(x) = 4x[1-4ik+3x(1+2n)], \nonumber \\
&&a_3(x) = -16x^3(1+2n)+4x^2[1+4ik+8n(n+1)]+ \nonumber \\ \nonumber && \qquad\qquad 4x(1+4k^2)(1+2n)
+2i(1-2k)(i+k)(i+4k),\\ \nonumber
&&a_4(x)= -32x^2(1+2n) +8x[-1+2n(n+1)+6ik]- \\ \nonumber && \qquad\qquad 4(i+k)(i+4k)(1+2n),\\
&&a_5(x) =4n(n+1)[4x(1+2n)+3(1-4ik)]. \nonumber
\end{eqnarray}
One can work out the indicial equation for this ODE (since there is
a regular singular point at $x=0$) and obtain

\begin{equation}
\sigma(\sigma-1)[\sigma^2-\sigma-4(1-k)(i+k)]=0
\end{equation}
where the solution to the ODE behaves as $x^{\sigma}$ as $x \to 0$.
The $\sigma=0$ solution of course corresponds to our polynomial
$\Lambda^k_n(x)$. Remarkably, one can write down the general
solution to this fourth order ODE, which is

\begin{eqnarray}
y(x)=c_1 I_{-1/2 +ik}(x) \, M_{n+1/2,ik}(2x) + c_2 I_{-1/2+ik}(x) \,
W_{n+1/2,ik}(2x) \nonumber \\+ c_3 K_{-1/2 +ik}(x) \, W_{n+1/2,ik}(2x) +c_4
K_{-1/2 +ik}(x) \, M_{n+1/2,ik}(2x).
\end{eqnarray}
Therefore, our polynomial must be such a linear combination and one
can use the asymptotics as $x \to \infty$ and $x \to 0$ to determine
all the constants uniquely. After a bit of work, one finds the
following:

\begin{eqnarray}
&&c_1=0, \qquad c_2 = 1+ c_4\frac{\pi \Gamma(1+2ik)}{\Gamma(-n+ik)}, \\
&&c_3 + \frac{2}{\pi}c_2 \cosh \pi k + c_4
\frac{\Gamma(-n-ik)}{\Gamma(-2ik)} =0, \\
&&2c_2 + \frac{\pi c_3}{\cosh \pi k}= \frac{\Gamma(-ik)\Gamma(1/2
+ik)\Gamma(-n+ik)}{\sqrt{\pi} \Gamma(2ik)\Gamma(-n-ik)},
\end{eqnarray}
which of course can be solved simultaneously. Before doing this let
us examine the $k =0$ limit. This will lead to $c_2=1$, $c_3=0$ and
$c_4= (-1)^{n+1} n!/ \pi$. Using~\cite{gr,aar}

\begin{eqnarray}
&&W_{n+1/2,0}(2x)= (-1)^n n! (2x)^{1/2}e^{-x}L^{0}_n(2x), \\
&&M_{n+1/2,0}(2x) = (2x)^{1/2}e^{-x}L^{0}_n(2x), \\
&&I_{-1/2}(x) = \sqrt{\frac{2}{\pi x}} \cosh x, \\
&&K_{1/2}(x)= \sqrt{\frac{\pi}{2x}} e^{-x},
\end{eqnarray}

one can show that
\begin{eqnarray}
\Lambda^0_n(x)= \frac{(-1)^n n!}{\sqrt{\pi}} L^{0}_n(2x),
\end{eqnarray}
as it should! Finally, solving for $c_2$, $c_3$ and $c_4$ we get,

\begin{eqnarray}
&&c_2 = 1- \frac{ik\Gamma(-ik)^2}{2^{2ik}\Gamma(2ik) \Gamma(-n-ik)^2},
\\
&&c_3 = -\frac{2}{\pi} \cosh \pi k + \frac{2ik \Gamma(-ik)^2 \cosh \pi
k}{2^{2ik} \pi \Gamma(-n-ik)^2}
+\\ && \qquad \nonumber \frac{\Gamma(-ik)\Gamma(-n+ik)}{\sqrt{\pi} \Gamma(2ik)\Gamma(1/2-ik)
\Gamma(-n-ik)}, \\
&&c_4 = - \frac{\Gamma(-ik)^2\Gamma(-n+ik)}{2\pi 2^{2ik} \Gamma(2ik)
\Gamma(-n-ik)^2}
\end{eqnarray}
which allows us to express the polynomial $\Lambda^k_n(x)$ in terms
of modified Bessel and confluent hypergeometric functions as
follows:

\begin{eqnarray}
\Lambda^k_n(x) = c_2 I_{-1/2+ik}(x) \,
W_{n+1/2,ik}(2x) + c_3 K_{-1/2 +ik}(x) \, W_{n+1/2,ik}(2x) \nonumber \\ +c_4
K_{-1/2 +ik}(x) \, M_{n+1/2,ik}(2x).
\end{eqnarray}
Now we will give an indication as how to derive the general solution
to the fourth order equation. Consider the coupled second order
equation (\ref{coupled}), and substitute the trial function
$\Lambda(x)=K_{-1/2+ik}(x)F(x)$. Using Bessel's equation to
eliminate second derivatives of $K_{-1/2+ik}(x)$ and the first order
recurrence relation (\ref{first}) to eliminate single derivatives of
$K_{-1/2+ik}(x)$, we find that the trial function satisfies
(\ref{coupled}) if

\begin{eqnarray}
L(F)=0, \qquad \textrm{and} \qquad F^*(x)=-F(x).
\end{eqnarray}
Thus possible choices for $F(x)$ are:

\begin{eqnarray}
&&iW_{n+1/2,ik}(2x), \nonumber \\ \nonumber
&&i(M_{n+1/2,ik}(2x)+ M_{n+1/2,-ik}(2x)) \nonumber \qquad \textrm{and} \\
&&M_{n+1/2,ik}(2x)- M_{n+1/2,-ik}(2x). \nonumber
\end{eqnarray}
Hence two solutions to the fourth-order equation are
$K_{-1/2+ik}(x)W_{n+1/2,ik}(2x)$ and
$K_{-1/2+ik}(x)M_{n+1/2,ik}(2x)$. To get the other two we need to
introduce $\tilde{I}_{-1/2+ik}(x)= I_{-1/2+ik}(x)+I_{1/2-ik}(x)$.
Notice that this function satisfies the same Bessel equation as
$K_{-1/2+ik}(x)$ does, and also a very similar first order
recurrence relation, namely $x\tilde{I}'_{-1/2+ik}(x)
-(-1/2+ik)\tilde{I}_{-1/2+ik}(x) =x \tilde{I}^*_{-1/2+ik}(x)$ (this
equation has a minus on the RHS for $K_{-1/2+ik}(x)$, see
(\ref{first})). Substituting the trial function
$\tilde{I}_{-1/2+ik}(x)G(x)$ leads to a similar condition on $G(x)$
as we obtained for $F(x)$:

\begin{eqnarray}
L(G)=0, \qquad \textrm{and} \qquad G^*(x)=G(x).
\end{eqnarray}
Thus $\tilde{I}_{-1/2+ik}(x)W_{n+1/2,ik}(2x)$ and
$\tilde{I}_{-1/2+ik}(x)M_{n+1/2,ik}(2x)$ solve the fourth-order
equation. Using the other two solutions, this means
${I}_{-1/2+ik}(x)W_{n+1/2,ik}(2x)$ and
${I}_{-1/2+ik}(x)M_{n+1/2,ik}(2x)$ are also solutions to the fourth
order equation, and hence we have completed the proof.

\section{Related identities}
We have derived an identity for $N=n+1/2$. There also exist similar
identities for $N=-n-1/2$. The case $N=\pm n$ is also of interest;
in this case the identities look like $W_{n,ik}(2x) =
\sqrt{x}p_n(x)K_{1+ik}(x)+ \sqrt{x}q_n(x)K_{ik}(x)$, where $p_n(x)$
and $q_n(x)$ are polynomials of degree $n$, but we shall not go
through the details here. Also note that similar identities probably
hold between $M_{N,ik}(2x)$ and $\tilde{I}_{-1/2+ik}(x)$, since the
function $\tilde{I}_{\nu}(x)$, like $K_{\nu}(x)$, is symmetric in
the order and hence
$\tilde{I}^*_{-1/2+ik}(x)=\tilde{I}_{1+(-1/2+ik)}(x)$, just like for
$K_{-1/2+ik}(x)$, which was an important property we used in the
proof.

\section*{Acknowledgments}
{ \small The author is supported by EPSRC. The author thanks
Sigbjorn Hervik for a useful conversation.}

\end{document}